# Effects of neutron irradiation on polycrystalline Mg$^{11}$B$_2$

C Tarantini[1], H U Aebersold[2], V Braccini[1], G Celentano[3], C Ferdeghini[1], V Ferrando[1], U Gambardella[4], F Gatti[5], E Lehmann[2], P Manfrinetti[6], D Marré[1], A Palenzona[6], I Pallecchi[1], I Sheikin[7], A S Siri[1] and M Putti[1]

[1] INFM-LAMIA/CNR, Dipartimento di Fisica, Via Dodecaneso 33, 16146 Genova, Italy
[2] Paul Scherrer Institut, Deptm.Spallation Neutron Source SINQ, CH-5232 Villigen, Switzerland
[3] Superconductivity Section, ENEA Research Center, Via E. Fermi 65, Frascati, 00044, Italy
[4] INFN Frascati National Laboratory, Via E. Fermi 40, Frascati, 00044, Italy
[5] Dipartimento di Fisica, Via Dodecaneso 33, 16146 Genova, Italy
[6] INFM-LAMIA/CNR, Dipartimento di Chimica e Chimica Industriale, Via Dodecaneso 31, 16146 Genova, Italy
[7] GHMFL, CNRS, 25 Avenue des Martyrs, BP 166, 38042 Grenoble, France

E-mail: tarantini@ge.infm.it

**Abstract.** We studied the influence of the disorder introduced in polycrystalline MgB$_2$ samples by neutron irradiation. To circumvent self shielding effects due to the strong interaction between thermal neutrons and $^{10}$B we employed isotopically enriched $^{11}$B which contains 40 times less $^{10}$B than natural B. The comparison of electrical and structural properties of different series of samples irradiated in different neutron sources, also using Cd shields, allowed us to conclude that, despite the low $^{10}$B content, the main damage mechanisms are caused by thermal neutrons, whereas fast neutrons play a minor role.
Irradiation leads to an improvement in both upper critical field and critical current density for an exposure level in the range $1\div2\cdot10^{18}$ cm$^{-2}$. With increasing fluence the superconducting properties are depressed. An in-depth analysis of the critical field and current density behaviour has been carried out to identify what scattering and pinning mechanisms come into play. Finally the correlation between some characteristic lengths and the transition widths is analysed.

## I. INTRODUCTION

Since its discovery, occurred in 2001[1], magnesium diboride has been one of the most widely studied superconductors. Although as early as in 1959 theoretical works predicted the multigap and multiband features[2] and in 1980 two-band superconductivity was clearly observed in doped SrTiO$_3$ by tunnelling measurements[3], only magnesium diboride presents a spectacular two gap behaviour in a favourable range of temperatures.

It was theoretically and experimentally demonstrated that the introduction of disorder can heavily modify the basic MgB$_2$ properties. It was predicted that interband scattering, induced by non-magnetic impurities, produces pair breaking and suppresses the critical temperature down to 20 K[4]; intraband scattering, instead, affects resistivity and leads to a change in the shape of upper critical field[5,6] which can exhibit a very different behaviour as compared to BCS predictions. Different approaches have been tried in order to increase upper critical field and critical current density without significantly changing the critical temperature with the main purpose of improving the properties of this material for technical applications. The usual way to do this is the chemical substitution on either Magnesium or Boron sites in order to affect selectively the two intraband electron diffusivities. Although $MgB_2$ has a very simple crystalline structure, it is not easy to dope; only a few experiments have been successful. Some of them, carried out in single crystals [7], wires[8] or thin films[9], have shown that Carbon doping in the Boron site significantly enhances $H_{c2}$ and at the same time it decreases the anisotropy. Aluminium substitution on Magnesium site either in bulk or single crystal samples gives a clear evidence of a gradual suppression of the upper critical field[10,11]. With both the substitutions critical temperature is suppressed, due to different possible causes: in fact it is necessary to take into account the different variations of the density of states (DOS) induced by different charge dopings; moreover, disorder introduced by doping might enhance interband scattering, also as an effect of structural deformations, and does not allow to carefully identify the real reasons for the changes in the superconducting properties. These problems can be overcome either by co-doping, as attempted with a substitution of Lithium and Aluminium on Mg sites by Monni *et al*[12] that, nevertheless, did not obtain a complete charge balance , or by particles irradiation.

With regard to particles irradiation, different kinds of experiments using α particles, heavy ions, protons or neutrons have been presented in the literature. The first two methods have been used on $MgB_2$ thin films. Damage by α particles has been carried out by Gandikota *et al.*[13,14], who found a gradual $T_c$ suppression and resistivity increase. 200 MeV Ag ions instead have been used by Shinde *et al.*[15], who observed that the critical temperature remains nearly unchanged, while there is a slight critical current improvement in a narrow range of magnetic fields. Concerning protons irradiation, both bulk and single crystals have been damaged[16,17]. In order to obtain a uniform defects distribution, it is necessary to vary the protons energy that determines the penetration depth. This method leads to an improvement of the $J_c$ field dependence as well as to a significant increase of the irreversibility field which is not accompanied by a significant $T_c$ suppression.

Neutron irradiation has been extensively studied and has shown to be very effective in introducing defects in $MgB_2$ which are able to improve the superconducting properties and also in suppressing the critical temperature strongly[18-22]. In ref.18 and 19 the annealing of defects produced by neutron irradiation has been investigated. Wilke[19] has irradiated $MgB_2$ wires with fluences ranging between $4.75 \cdot 10^{18}$ and $1.9 \cdot 10^{19}$ cm$^{-2}$, obtaining a critical temperature lower than 5 K even in the least irradiated samples. Then, the superconductive properties have been systematically studied as a function of annealing temperature and time; using this procedure only slight $H_{c2}$ and $J_c$ improvement with respect to those of the pristine samples has been observed.

In general neutron irradiation suffers of the inhomogeneity problem due to the huge cross section of neutron capture by $^{10}B$ whose consequence is the short penetration depth of thermal neutrons in $MgB_2$. In ref.20 and 21 this problem has been overcome by using a Cadmium shield to absorb the lowest energy neutrons.

Another method to avoid inhomogeneous defects distribution in bulk and wires is the choice of sample thickness smaller than the penetration depth[18,19].

In this work we solve the inhomogeneity problem, inherent in neutrons irradiation, by preparing samples with boron isotopically enriched in $^{11}B$, so as to reduce the $^{10}B$ capture probability and to enlarge the penetration depth, following the idea presented in ref.22. We present a wide investigation of superconducting properties for a large series of samples, and we compare our results with those of irradiated samples of literature prepared with natural boron.

## II. SAMPLE PREPARATION AND CHARACTERIZATION

The samples for this experiment were prepared by direct synthesis of pure elements with a single step method[23], using a similar technique as in earlier works[24,25]. Boron and Magnesium (99.999% purity) were put in Ta crucibles welded in Argon and closed in a quartz tube under vacuum. Then they were heated up to 950° C to produce dense, clean and hard cylinder shaped samples. The only peculiarity of these samples, compared to the usually prepared ones, comes from the particular boron used. In fact we employed crystalline isotopically enriched $^{11}B$ from Eagle-Picher (99.95% purity), with a residual $^{10}B$ concentration lower than 0.5%. The pristine properties are optimal $T_c$ (39.2 K), sharp superconducting transition (0.2 K), low residual resistivity ($\rho(40) < 2\mu\Omega cm$) and high residual resistivity ratio (RRR ~ 11÷15). The quality of the $MgB_2$ phase was initially studied by X-ray powder diffraction with a Guiner-Stoe camera; no extra peaks due to free Mg or spurious phases were detected. The samples were then cut in parallelepiped bars (~1x1x12 mm$^3$) and irradiated in different ways.

### A. Neutron irradiation

In order to shed light on the different role of fast and thermal neutrons, three series of samples were irradiated at two different facilities. The first series (L-series) was irradiated in a TRIGA MARK II type nuclear research reactor (thermal and fast neutron flux density up to $10^{13}$ cm$^{-2}$s$^{-1}$ and $1.2\cdot10^{13}$ cm$^{-2}$s$^{-1}$, respectively) at the Laboratory of Applied Nuclear Energy (LENA) of the University of Pavia, reaching a maximum thermal neutron fluence of $10^{18}$cm$^{-2}$. The other two series were irradiated in a spallation neutron source SINQ (thermal and fast neutron flux density up to $1.6\cdot10^{13}$ cm$^{-2}$s$^{-1}$ and $10^{10}$ cm$^{-2}$s$^{-1}$, respectively) at the Paul Scherrer Institut (PSI), Villigen. In this latter facility, we exposed $MgB_2$ samples to neutrons, both with and without Cd shields (P-Cd and P-series respectively), for increasing irradiation times, to gradually modify the superconducting properties. To obtain an exposition level of $1.4\cdot10^{20}$cm$^{-2}$, the most irradiated sample was exposed for nearly four months. It is worth noticing that these two facilities have approximately the same flux density of thermal neutron, but they differ by three orders of magnitude in the fast neutron flux density.

Fast or thermal neutrons should induce different types of defects as a consequence of their interaction reactions and the corresponding cross-sections, which are dependent on the neutron energy. Fast neutrons have large enough energy to undergo direct collisions with nuclei and displace them from their lattice positions, thus producing point defects. However, the cross-sections for such events are generally smaller than the ones for thermal neutrons; this means that, in general, thin samples are not strongly influenced by fast neutrons. Thermal neutrons instead have lower energy and the only way they produce disorder is through the neutron capture reaction by $^{10}$B: this process causes the isotropic emission of an $\alpha$ particle and a $^{7}$Li nucleus, with energy of 1.47 MeV and 0.84 MeV, respectively. Along the range of recoil (4.8 µm and 2.1 µm, respectively) these particles loose energy by interacting with electrons and they create atom displacements mainly at the end of the range. As this capture reaction has a large cross section, if we used natural boron (with 20% of $^{10}$B content), the penetration depth would be as small as 200 µm and there would be a shielding effect at the surface. In thicker bars, this should give rise to a much disordered region next to the surface and therefore a gradient in the defects density throughout the sample. We chose to employ isotopically enriched $^{11}$B to avoid this problem; in fact with only 0.5% of $^{10}$B content, the penetration depth grows up to ~1 cm, becoming larger than our MgB$_2$ bars thickness.

### B. Structural characterization

The general properties of the so obtained samples are summarized in table 1, where critical temperature, resistive transition width (10÷90% criterion), normal state resistivity, $\Delta\rho=\rho(300K)-\rho(40)$, residual resistivity ratio RRR=$\rho(300K)/\rho(40)$ and crystallographic axes at different thermal neutron fluence are reported. In table 1, we report also the data for the Cadmium shielded sample: this results will be separately described in a following paragraph.

X-ray diffraction, performed in the standard Bragg-Brentano geometry, allowed us to study the structural properties before and after irradiation. Figure 1 shows the magnification of the (002) and (110) MgB$_2$ reflections for some irradiated samples of the PSI series. The peaks remain very narrow up to the highest exposition level and the gradual changes in both lattice parameters are evident. The double peak structure is due to CuK$_{\alpha1}$ and CuK$_{\alpha2}$ present in the X-ray beam. In order to carefully estimate *a*- and *c*-axes, we performed a Rietveld refinement on the whole X-ray spectra for all the samples. In figure 2 we report the lattice parameters as a function of the thermal neutron fluence of the sources. The lattice parameters increase following the same trend in both L- and P-series; the *a* axis expands by less than 0.4%, whereas the *c* axis increases by more than 1% (volume change of 1.7%). A similar anisotropic expansion has been observed also by Kar'kin *et al.*[18], who reported *a* and *c* variations of 0.24 and 0.9%, respectively, with a volume increase of 1.4% at $10^{19}$ cm$^{-2}$ fluence level, comparable to our value 1.1% at the same irradiation. The behaviour presented by Wilke *et al.*[19] is quite different. At just 9.5·$10^{18}$ cm$^{-2}$, they found an enhancement of volume by 2.6%, greater than the values we obtain in the most irradiated samples. Moreover the authors showed a substantial change in the structural properties of the samples upon irradiation with a strong (002) peak broadening; this trend, observed on the most irradiated samples, is attributed to a change in the correlation length along the *c* axis. Instead, in the less irradiated wires the peak widths were similar to ours. The (002) peak broadening probably appears at a damage level not yet reached in our samples, as a consequence of our low content of $^{10}$B.

## C. Electrical characterization

Resistivity and magnetoresistivity were investigated by AC electrical resistance measurements carried out in a 9 T Quantum Design PPMS and then up to 28 T in a resistive magnet at the Grenoble High Magnetic Field Laboratory with a measuring current density of ~1 A/cm$^2$. The resistivity curves are plotted in figure 3 for P-series; moreover in the inset the susceptibility, measured by Quantum Design SQUID magnetometer in an applied field of 10 Gauss, is shown. The transition widths, measured both by resistivity and susceptibility, remain rather sharp even in the most irradiated samples. Only sample P-4, in the magnetic transition, shows a width larger than 3 K. The overall transition width versus fluence will be discussed in par.VI.

In figure 4 we plot the critical temperature $T_c$ (fig.4a) (defined at 50% of normal state resistivity) and the resistivity ρ at 40 K (fig.4b) as a function of the thermal neutron fluence. The critical temperature of the P- and L-series decreases monotonically with the neutron fluence: it is suppressed only by 2 K up to $10^{18}$ cm$^{-2}$ and then reaches 9.1 K at the highest exposition level. At the same time resistivity grows by more than two orders of magnitude, from ~1 to 130 μΩcm. The resistivity increase is likely due to an enhancement in intraband scattering, induced by irradiation, which causes a reduction in the electron mean free path. With regard to Δρ, it remains approximately constant (~10÷15 μΩcm) over the whole range of irradiation (see table 1). In ref.26 Rowell suggested that in poly-crystalline samples Δρ should be nearly equal to 8÷9 μΩcm, while an increasing Δρ in irradiated and substituted samples indicates poor connectivity. In the pristine samples (LENA-0 and PSI-0) we measure generally Δρ values slightly larger (10-12 μΩcm) than the predicted ones, probably due to the not full density of the samples. Anyway, no systematic increase of Δρ with irradiation is observed, indicating that the produced damages do not significantly affect the connectivity of the samples[13].

To highlight the direct relationship between ρ and $T_c$ and eliminate the fluence dependence, in the inset of figure 5 we plot critical temperature as a function of resistivity (solid symbols). We observe a linear decrease of $T_c$ versus ρ, as already found in MgB$_2$ thin films irradiated by α particles[13]. Even if we rescale the resistivity data by assuming Δρ = 8 μΩcm as suggested by Rowell[26], we obtain again a linear behaviour (inset of figure 5, open symbols). With this correction, we obtain the limit resistivity value, that is 110 μΩcm, at which $T_c$ = 0; this resistivity is similar to value obtained in ref.13. In the main panel of figure 5, we compare our results on $T_c$ versus ρ (black line) with other similar data from literature about samples prepared with natural boron; we considered irradiated and then annealed thin films[13,14] and wires[19], irradiated bulk[27] and C-doped wires[28,29]. Almost all these series of samples follow a roughly linear $T_c$ versus ρ behaviour with a limit resistivity value ranging between 70 and 120 μΩcm. In particular, in Gandikota's thin films this behaviour is followed both before and after annealing procedure[14]. The only exception is observed in irradiated and successively annealed wires[19] that show very low resistivity values in spite of the large $T_c$ suppression. Interestingly, $T_c$ versus ρ behaviour of C-doped wires[28,29] nearly overlaps data of irradiated samples. As shown in par.IV these samples present nearly the same upper critical fields of irradiated ones.

It is worth noting that, differently by the Golubov and Mazin predictions[4], no saturation is present at 20 K: this means that pair breaking by interband scattering is not the only mechanism that is able to suppress the critical temperature in the irradiated samples. On the other hand, the roughly linear trend of $T_c$ with $\rho$ suggests that a strong correlation exists between the mechanism increasing the resistivity (scattering by atomic scale defects) and those which suppress the critical temperature. This behaviour is common to other superconductors. In amorphous transition metals and damaged A15 superconductors, for instance, a smearing of the peak in the electron density of states at the Fermi level was proposed as a mechanism for the $T_c$ reduction[30]; yet, this mechanism cannot simply explain the $MgB_2$ case, whose density of states is rather flat around the Fermi level; other mechanisms suppressing the electron-phonon coupling should be invoked as well.

## III. THERMAL OR FAST NEUTRONS

In the last years several experiments of neutron irradiation have been carried out to understand the effectiveness of thermal and fast neutrons. Nevertheless, up to now, the comprehension of the different effects coming into play is not clear, yet[19,20,21].

Now, we point out that both figure 2 and 4 are plotted as a function of the thermal neutron fluence. Although the two different neutron sources employed to irradiate the samples differ by three order of magnitude in the fast neutrons flux, both the L- and P-series follow the same behaviour both as for the crystal axes expansion and the change of critical temperature and resistivity. This is an evidence that the superconducting properties are effectively changed by thermal neutrons, whereas fast neutrons play a minor role.

To confirm this result we irradiated samples employing a Cadmium shield as proposed by Eisterer[20]. As Cadmium has a huge cross section for neutrons whose energy is lower than 0.5eV, this technique can be used to probe the fast neutrons role alone. The properties of such samples have been measured and analyzed as L- and P-series. The behaviour of these samples is quite different from the ones irradiated without shield. Already in the structural characterization (see fig.2) we note that lattice parameters of P-Cd1 have no changes compared to the pristine ones. P-Cd2 axes instead only slightly increase. The critical temperature and the resistivity values remain nearly unchanged. We can observe in table 1 and figure 4 that $T_c$ is suppressed only by 0.2 K in the most irradiated samples, whereas $\rho$ assumes approximately the same values of L-3 and P-1. These data indicate that this series behaves as if it was irradiated with fluences which are nearly 50 times smaller than those used for the L- and P-series. This effect can be easily explained considering that Cadmium shields thermal neutrons and only fast neutrons can reach the samples. We think that this finding confirms that fast neutrons are not very effective in the defect production. Only thermal neutrons and their nuclear reaction with residual $^{10}B$ should be taken into account to explain the modification of superconducting properties.

## IV. UPPER CRITICAL FIELD

Upper critical field is one of the more interesting properties and, in different experiment, the effect of the irradiation was studied on samples prepared with natural boron both with and without Cadmium shield[20,21]. It was observed that the presence of Cd-shield leads to a critical temperature close to the pristine value and an upper critical field only slightly enhanced in the parallel direction. Moreover, Eisterer[21] reported an evident anisotropy reduction in single crystals, induced mainly by an enhancement of $H_{c2}$ in the perpendicular direction. In addition, that work reported also the study on the samples irradiated without Cadmium shield: $T_c$ and consequently also $H_{c2}$ were strongly suppressed, without substantially changing the $H_{c2}(T)$ shape. Wilke *et al.*[19], indeed, irradiated the samples up to $1.9 \cdot 10^{19} cm^{-2}$, which leads to a suppression of the critical temperature down to 5 K. To restore the superconducting properties, they annealed the samples at different temperatures; in the best case, when the critical temperature was almost completely restored, the upper critical field at 0 K grew from ~16 T in the undamaged sample, to about 19 T. Moreover, the curve shape slightly changed, showing a more marked upward curvature near $T_c$ and assuming a more linear trend at low temperature.

In this work, upper critical field has been evaluated for each temperature as the 90% of the normal state in the resistivity transition. It is worth noting that in polycrystalline samples, with randomly oriented grains, the so estimated values coincide with the largest $H_{c2}$, that is the critical field parallel to the *ab*-plane. In figure 6 we report $H_{c2}(T)$ curves; for clarity only P-series is shown. $H_{c2}$ curves of Cd-shielded one have only a slight improvement compared to pristine sample, while L-4 behaves similarly to P-2 curve. Moreover, in our data, the curves are gradually modified by neutron irradiation and they eventually assume a completely different shape. Initially, at low temperature, a remarkable downward curvature is observed. Upon irradiation, the slope increases and the behaviour becomes almost linear in P3.5 sample. Here, $H_{c2}(0)$ attains value larger than 30 T, twice as much as that of the pristine sample. With further irradiation, the curves show an upward curvature at low temperature, particularly evident in P3.7 and P4 samples. $H_{c2}(T)$ curves, in samples P5 and P6, are linear in the measured range of temperatures; the extrapolated values at 0 K turn out similar to ones shown in ref.14, 19. In the inset, the critical field at $T_c/2$ versus fluence for the P-samples series is plotted. In this graph the effect of irradiation on $H_{c2}$ is emphasized. $H_{c2}$ grows up to a fluence level of $2 \cdot 10^{18}$ cm$^{-2}$ (sample P-3.5), where the critical temperature is suppressed by only few degrees; instead, in the most irradiated samples, $H_{c2}$ is suddenly reduced both as a consequence of a strong $T_c$ suppression and of the anisotropy reduction observed in ref.21.

As a substantial improvement of upper critical field upon irradiation has been observed, like in Carbon doped samples, a comparison between our $H_{c2}$ data and those reported in literature on other disordered samples, could be interesting. In figure 7 we report the upper critical field at 0 K, obtained by linear extrapolation, as a function of critical temperature for different literature samples. We considered samples whose critical temperatures and resistivities of the starting materials were similar to ours: thin films irradiated with α–particle[14], Carbon substituted wires[28], neutron irradiated bulks[27] and wires[19]. All these disordered samples behave similarly: upper critical field presents a maximum, larger then 35 T, when critical temperature is around 35 K. This behaviour was already observed in earlier works[27,22,14,28] and underlined in ref.19 and 31. The behaviour of samples annealed after irradiation is quite different (full symbols); in fact, both in Gandikota's thin films[14] and in Wilke's wires[19], the upper critical field remains significantly lower. It appears that annealing is more effective in removing the defects related to the increase of $H_{c2}$.

In order to understand the $H_{c2}$ behaviour, we applied the theoretical model of ref.5 and 9, where the Usadel equations for a dirty two-gap anisotropic superconductor are developed. Starting from these equations, the behaviour of the upper critical field as a function of temperature is derived. These articles emphasize as interband scattering should be the principal cause for the critical temperature suppression, whereas intraband scattering, and thereby diffusivities ratio ($\eta = D_\pi/D_\sigma$), affects the shape of $H_{c2}$ vs T curves in different temperature ranges. If $\eta > 1$, the critical field shows a slight upward curvature near $T_c$, while at 0 K it exceeds the predicted BCS value; oppositely, if $\eta < 1$, $H_{c2}$ remains nearly linear in the high temperature region and presents a remarkable upward curvature at low temperature. This model seems to explain the upper critical field behaviour[9,32,33,34], even if it does not allow to estimate the resistivity values correctly. If we fix the electron-phonon coupling constants of the pure $MgB_2$, we can reproduce the shape of the less irradiated samples, as long as $T_c$ is not reduced by more than 6 K and critical field is steadily increased by irradiation. In table 2 the fit parameters for P1, P2, P3 and P3.5 are summarized; since in bulk samples it is possible to measure the critical field in one direction only, we could solely determine the effective σ-band diffusivity $D_{\sigma Eff}=(D_\sigma^{ab}D_\sigma^{c})^{0.5}$ and the effective ratio $\eta_{Eff}=D_\pi/D_{\sigma Eff}$ ($D_\pi$ is assumed isotropic). Nevertheless we have an estimation of the anisotropy values from the critical current analysis[35] (as mentioned later on in section V, γ ranges from 4.5÷4, in the less irradiated samples, to 1, in the most irradiated ones), so we can determine both in-plane and out of plane σ-band diffusivities. In all the fits, the critical temperature values are smaller than the experimental ones by 1 or 2 K; this problem does not worsen the fit quality as shown in figure 8, where calculated curves are plotted together with data points for P-3 and P-3.5. It turns out that $\eta = D_\pi/D_\sigma^{ab}$ is larger than unity and monotonically decreases from 9.5 to 5.7 with increasing irradiation (see table 2); in the framework of this model, these η values indicate that σ-band is always dirtier than π-band. Moreover $D_\sigma^{ab}$ and $D_\pi$ are approximately reduced by a factor 3 and 6, respectively, from P1 to P3.5; this means that π-band gets dirty faster than σ-band. From $D_\pi$ and $D_\sigma$ we can estimate the residual resistivity; the obtained values, reported in table 2, increase with irradiation as in the measured ones, but they are slightly larger.

However $H_{c2}$ model does not work in the most irradiated samples, perhaps due to a change in the electron-phonon coupling. Such hypothesis seems to be confirmed by the specific heat studies carried out on these samples[36]; we have noticed that Sommerfeld's constant $\gamma_S$ remains unchanged (~3 mJ/mol·K$^2$) in the less irradiated samples, and then decreases to 2.5 mJ/mol·K$^2$ in the most irradiated ones. As $\gamma_S$ linearly depends on the density of states renormalized by the electron-phonon coupling, the decrease of $\gamma_S$ can be attributed to a weakening of the coupling. This effect, besides being an explanation for the $T_c$ suppression, can be accountable for the failure of the $H_{c2}$ model. As in the most irradiated samples the coupling constants change, we should take into account this modification to fit the $H_{c2}$ curves. As mentioned above, P3.7 sample shows a marked upward curvature in $H_{c2}$ in the whole range of temperatures and this trend cannot be explained by this model. In a sample whose critical temperature is so much suppressed indeed, also in the case of $\eta < 1$ the upward curvature should not appear, as a consequence of a large interband scattering. Probably this critical temperature is mainly determined by other mechanisms, rather than by the interband scattering. Instead, in the most irradiated samples, critical temperature is low and no upper curvature is observed near $T_c$. In ref. 19 a WHH single gap behaviour[37] has been observed in the case of similar critical temperature. In our less irradiated samples, $H_{c2}$ measurements are limited to about $T_c/2$ but the presence of a single gap is supported by specific heat results. In fact they highlight a crossover between two and single gap behaviour when critical temperature passes from 21 to 11 K[36].

The diffusivity parameters that we obtained in the framework of this $H_{c2}$ model produce acceptable fits, at least on the less disordered samples. Nevertheless some cautions must be taken in considering the diffusivity values. In fact the calculated resistivities, even though they show the same monotonic trend with respect to the measured ones (ρ increases with disorder), do not correspond exactly as for their absolute values. In ref. 38 the authors suggested that this discrepancy could be induced by an additional extrinsic contribution in the measured resistivity due to grain boundary structure. However, this interpretation cannot be applied in our case because the resistivities estimated by the fit parameters are larger than experimental ones. Moreover, despite it seems reasonable that increasing irradiation the two bands tend to become equally disordered (η → 1), because the disorder should not be preferentially located in the Boron or Magnesium planes, the fact that, in the less disordered sample, σ-band turns out to be dirtier than π-band is a controversial result. In literature, in general, in clean systems σ-band is believed to be cleaner than π-band [39], although there are also opposite results[34]. We believe that the above mentioned results need to be confirmed by similar analysis done on critical fields of single crystal or clean epitaxial films where the presence of the two orientations reduces the number of free fit parameters. Recently we demonstrated that high field magnetoresistance[40] is an alternative method to determine the scattering times in the two bands: work is in progress to perform similar measurements (that need high magnetic fields) on these samples and study the critical fields of irradiated epitaxial thin films.

## V. CRITICAL CURRENT DENSITY

In order to examine the effect of neutron irradiation, the critical current density and its behaviour in magnetic field have also been analysed. The hysteresis loops have been measured in a Quantum Design SQUID magnetometer up to 5 T and in a Vibrating Sample Magnetometer (VSM) up to 12 T. The critical current densities were extracted applying the critical state model. In figure 9, we report experimental data of critical current density of all P-samples, some of which have been studied in a previous publication[35]. Irradiation, despite suppressing $T_c$, significantly improves the magnetic field dependence of the critical current density. In the inset of figure 9, the trend of $J_c$ versus neutrons fluence at 5 K and 4 T is plotted. Again, like in the $H_{c2}$ behaviour described above, this is an indication that optimised superconducting properties are obtained by irradiating with fluence near $2 \cdot 10^{18}$ cm$^{-2}$, close to the case of the P3.5 sample. This sample has a critical current smaller than P3 up to 4 Tesla; however at this point a crossover occurs and P3.5 shows a strong improvement of its field dependence, reaching a value close to $10^4$ A/cm$^2$ at 10 Tesla.

It is evident that the rise of $J_c$ with irradiation cannot be ascribed to the increase of $H_{c2}$: at low irradiation levels ($10^{17}$ cm$^{-2}$), despite the fact that the upper critical field does not change significantly, the field dependence of $J_c$ is strongly affected by irradiation. Indeed, by applying Eisterer's percolative model[41], we have demonstrated quantitatively that neutron irradiation induces a new pinning mechanism by point-like defects, which is responsible for the significant improvement of the critical current behaviour in magnetic field[35]. We assumed that in the pristine sample the dominant pinning mechanism is by grain boundaries and that such contribution is unaffected by irradiation, so that it should rescale with the condensation energy of the sample $E_c = \mu_0 B_c^2(T) \approx \mu_0 B_c^2(0)(1-(T/T_c)^2)^2 \propto T_c^2(1-(T/T_c)^2)^2$. The rescaled grain boundary pinning contribution lies well below the experimental data for samples irradiated with fluence lower than $10^{18}$ cm$^{-2}$, that is when the critical temperature is not depressed by more than 10%. The total current can be calculated as the sum of the rescaled grain boundary contribution plus a point defect contribution, with two free parameters: the point defects pinning multiplicative coefficient $A_p$ and the anisotropy γ. The fitting results show that the anisotropy decreases with increasing fluence, with values around 4-4.5 for samples irradiated at fluences up to $10^{18}$ cm$^{-2}$ (from P0 to P3.5), and around 1 for samples irradiated at larger fluences (from P4 to P6), while the coefficient $A_p$ increases with fluence. In particular, $A_p$ values scale with the 1/3 power of the fluence; this is a check of the validity of this description, as $A_p$ is indeed linearly related to the inverse of the average distance between point defect pinning centers induced by irradiation, which in turns is proportional to the 1/3 power of the fluence. Instead, as confirmed also by new data on P3.5 and P3.7 samples, for fluences larger than $10^{18}$ cm$^{-2}$ the rescaled grain boundary contribution is even larger than the experimental data points, indicating a failure of the scaling procedure. Work is in progress to understand the reason for such failure; we propose that it may be ascribed to two effects that we have neglected: the first one is the likely suppression in the density of states and the second one is the trend towards a single gap behaviour in the most irradiated samples, as demonstrated in ref. 36.

By a close inspection of the shape of the curves in the main panel of figure 9, it can be noted that while for P0, P1, P2, the critical current density monotonically decreases with magnetic field, in the cases of P3, P3.5, P3.7 it presents a smooth plateau/upraise (well visible in linear scale). At larger fluences this feature progressively disappears. A similar, but more evident, upraise has been observed on irradiated single crystals[42] where the peak position is shifted to lower magnetic fields with increasing disorder; this fishtail effect has been attributed to an order-disorder transition of the flux line lattice. Also pinning by variations of the superconducting order parameter, either due to areas with depressed $T_c$ or with depressed electron mean free path could result in a non-monotonic behaviour of the critical current density as a function of the magnetic field. Indeed, it may be that beside the already cited point-like defects pinning mechanism, also a pinning mechanism by local variations of the superconducting order parameter could be induced by neutron irradiation. Such other mechanism should be considered to fit the whole shape of the curves; however the large number of free parameters makes unreliable any quantitative analysis, especially due to the fact that the original pinning contribution by grain boundaries of the unirradiated sample does not scale as expected with $T_c$ for fluences larger than $10^{18}$ cm$^{-2}$. The single grain contribution to the critical current density $j_c^{(sg)}$ of the three different pinning mechanisms, namely by grain boundaries, by point defects and by variations of the superconducting order parameter, is shown in figure 10, as a function of the reduced field $H/H_{c2}$. It is evident that the presence of the latter mechanism could indeed account for the non monotonic magnetic field dependence of the sample critical current density, which results from the percolation process through randomly oriented grains.

## VI. CHARACTERISTIC LENGTHS

Some characteristic lengths, namely the coherence length and the electron mean free path, can be estimated in the irradiated samples in order to shed light on the type of disorder introduced by irradiation.

We have noticed that, with the exception of the intermediate range of fluence, the transition widths of these samples remain extraordinarily sharp despite the critical temperature is reduced by more than four times compared to the pristine value. This result is even more surprising as they have been observed also in the specific heat measurements[36] which are sensitive to all kinds of inhomogeneity. Moreover, it is interesting to note that P-3.7 and P-4 show a $\Delta T_c$ slightly larger than other samples and this is even true when $\Delta T_c$ is estimated by susceptibility and specific heat[36].

The sharpness of transition is indicative that defects induced by irradiation are homogeneously distributed on the scale of the coherence length. The evaluation of the coherence length $\xi$ cannot be obtained straightforwardly from $H_{c2}$ because of two band contributions. Anyway, we can estimate an effective value of $\xi$ by the relation $\xi = \left[ \dfrac{2\pi\Phi_0}{\mu_0 H_{c2}^{//c}(0)} \right]^{1/2}$, where $\Phi_0$ is quantum flux ($2.07 \times 10^{-15}$ T·m$^2$) and $\mu_0 H_{c2}^{//c}(0)$ is the upper critical field parallel to the *c*-axis at zero temperature. As already said, in polycrystalline samples we measure the uppermost critical field, i.e. $\mu_0 H_{c2}^{//ab}(0)$, and $\mu_0 H_{c2}^{//c}(0)$ can be estimated as $\mu_0 H_{c2}^{//ab}(0)/\gamma$ where $\gamma$ comes from $J_c$ analysis[35]. The so derived $\xi$ values are shown in figure 11 as a function of fluence. In spite of several approximations the result is clear, reflecting the behaviour of the upper critical fields. Starting from the clean sample ($\xi \sim 100$ Å), the coherence length decreases by increasing the disorder, as an effect of the suppression of the mean free path. A further increase of disorder suppresses $T_c$ and therefore the condensation energy. This causes an enhancement of $\xi$, which reaches again a value of about 100 Å in the most irradiated sample.

The amplitude of the resistive transition, $\Delta T_c = T_{90\%} - T_{10\%}$, is plotted in the same graph (black circles); $\Delta T_c$ and $\xi$ curves show a mirror like behaviour. The transition width $\Delta T_c$ reaches its maximum, about 1 K, in P-3.7 sample and, in the same sample, the coherence length has a minimum (~ 40 Å). In the most irradiated samples the transition is sharp again ($\Delta T_c \sim 0.3$ K), being the coherence length raised to 100 Å. The strong correlation existing between $\Delta T_c$ and $\xi$ indicates that when coherence length is much larger than the typical scale of defect inhomogeneity, the disorder is averaged out; on the contrary when the coherence length decreases (P-3.5, P-3.7 and P-4) the disorder is more effective in broadening the transition. Consistently with this scenario, the samples with smaller coherence lengths are the only ones whose critical current exhibits a broad bump characteristic of pinning by local inhomogeneity of the superconducting order parameter (see section V). To verify that transition width behaviour is not influenced by the employed criterion, in figure 11 we add also data obtained by 2÷98% of residual resistivity (grey stars); we note that the same trend is shown.

Another length that characterizes disordered samples is the electronic mean free path, $l$. In a two band system two different mean free paths, $l_\sigma$ and $l_\pi$, should be considered. If we are interested only to emphasize the behaviour of $l$ with disorder, an effective value can be extracted. We suppose, for simplicity, that the scattering rates in the two bands are the same ($\Gamma_\pi \sim \Gamma_\sigma = \Gamma$); this assumption is quite reasonable in artificial disordered samples. The scattering rate can be evaluated by resistivity as $\Gamma = \varepsilon_0 \rho (\omega_{P\sigma}^2 + \omega_{P\pi}^2)$, where $\omega_{P\sigma} = 6.23$ eV and $\omega_{P\pi} = 3.40$ eV are the mean values of the plasma frequencies[43]. By the mean in-plane value of Fermi velocity ($v_{F\sigma} \sim v_{F\pi} \sim 5.1 \cdot 10^5$ m/s) we determine the mean free path $<l^{ab}>$. $<l^{ab}>$ (full symbols) is plotted as a function of corrected residual resistivity in figure 12. The initial value, 600 Å, indicates a very clean sample condition; then $<l^{ab}>$ decreases down to few Å, the same order of magnitude of the cell size.

Finally, it could be interesting to compare the mean free path with the BCS coherence length ($\xi_0$) in order to verify the passage from the clean to the dirty limit in irradiated samples. The two band nature of MgB$_2$ does not allow an unambiguous definition of dirty limit, even if it was proved that the increase of the upper critical field with respect to its intrinsic value occurs when σ-bands become dirty ($l_\sigma < \xi_{0\sigma}$)[10]. As discussed above, we are not able to estimate $l_\sigma$, while $\xi_{0\sigma}$ can be calculated from the energy gap of σ-bands, $\Delta_\sigma(0)$, determined in ref.36 by employing the formula $\xi_{0\sigma} = \frac{\hbar v_{F\sigma}}{\pi \Delta_\sigma(0)}$. In figure 12 we report $\xi_{0\sigma}$ (open symbols): it ranges from 150 Å, in the pristine sample, to nearly 900 Å, in the most exposed one. In the limit of the rough approximations done, we can observe that in the pristine sample $\xi_{0\sigma}$ is close to the $\xi$ estimated by the upper critical field. This confirms that unirradiated samples are in the clean limit[10]. At fluence larger than $1.0 \cdot 10^{17}$ cm$^{-2}$, H$_{c2}$ increases and consequently $\xi$ diminishes; this behaviour implies the crossover from the clean to the dirty limit. In the dirty limit the coherence length $\xi$ is related with the mean free path $l$ and the BCS coherence length $\xi_0$ by the relation $\xi \approx (\xi_0 l)^{1/2}$. Thus, looking at figure 11, the non monotonic behaviour of $\xi$ can be nicely explained by the decrease of $l$ at low level of disorder, followed by the sharp increase of $\xi_{0\sigma}$ at high level of disorder.

## VII. CONCLUSION

We prepared three different series of high quality polycrystalline Magnesium Diboride by direct synthesis of pure elements. To avoid self shielding, in the sample preparation we used Boron with less than 0.5% $^{10}$B, thus increasing penetration depth of neutrons up to 1 cm. The samples were irradiated with neutrons at two different facilities that have approximately the same thermal neutron flux density, but differing by three orders of magnitude in the fast neutron flux density. A series of samples was irradiated with Cd shield to absorb thermal neutrons. The comparison of electrical and structural properties of the three sample series allowed us to conclude that the main damage mechanisms are caused by thermal neutrons, whereas fast neutrons play a minor role.

The critical temperature was reduced down to 9 K with a fluence of $1.4 \cdot 10^{20}$ cm$^{-2}$; at the same time the normal state resistivity is increased by more than two orders of magnitude and the cell volume is considerably increased (1.7%). All the samples, even the most irradiated ones, showed a good level of homogeneity. This is proved by X-ray diffraction patterns, whose peaks remain very narrow up to the highest exposition level. Moreover, both in resistive and inductive measurements, the transition widths remain sharp, showing a broadening only in the intermediate range of fluence.

The upper critical field appears strongly influenced by irradiation. At low fluences $H_{c2}$ increases exceeding 30 T at $2 \cdot 10^{18}$ cm$^{-2}$, while the critical temperature is only slightly reduced. For higher fluences $H_{c2}$ is suddenly reduced both as a consequence of a strong $T_c$ suppression and the anisotropy reduction. We compared these data with the current literature ones and we observed a pronounced maximum in $H_{c2}$ both in irradiated and C-doped samples at about 35 K of $T_c$.

We interpreted $H_{c2}$ data in the framework of an existing model for two bands superconductors: despite the fit quality is acceptable, the resistivity calculated from the obtained scattering times does not match perfectly with the measured one. The interband scattering might not be the only mechanism able to suppress superconductivity. Another likely mechanism is the weakening of electron-phonon coupling, as supported by specific heat measurements.

The critical current density as well undergoes the effect of neutron irradiation showing a nearly flat magnetic field dependence in a region as wide as several Teslas. The largest $J_c$ improvement was obtained at the same exposition level, where the largest increase of the upper critical field was observed. We had clear evidence that, to take into account the evolution of $J_c$ with fluence, it is necessary to include the contributions of more pinning mechanisms: pinning by grains boundaries, by point-like defects and by local variations of the superconducting order parameter. The relative weight of these contributions changes as a function of fluence. Finally a correlation between coherence length and the transition width has been shown: we evaluated that in the samples with the larger coherence lengths the transition appears extremely sharp because disorder is averaged out, whereas in the samples with lower coherence lengths disorder is more effective in broadening the transition. Consistently with this scenario, samples with smaller coherence lengths are those whose critical current exhibits a broad bump characteristic of pinning by local inhomogeneity of the superconducting order parameter.

**Acknowledgments**

We acknowledge the support of the European Commission from the 6th framework programme "Transnational Access - Specific Support Action", contract N° RITA-CT-2003-505474". This work is partially supported by the project PRIN2004

**Table 1.** Main properties for all the irradiated samples: thermal neutron fluence, critical temperature and transition width (defined as $T_c = T_{50\%}$ and $\Delta T_c = (T_{90\%} - T_{10\%})$), resistivity at 40 K, $\rho(40\ K)$, residual resistivity ratio RRR=$\rho(300K)/\rho(40K)$, lattice parameters *a* and *c*.

| Samples | Fluence (cm$^{-2}$) | $T_c$ (K) | $\Delta T_c$ (K) | $\rho(40)$ ($\mu\Omega$cm) | $\Delta\rho$ ($\mu\Omega$cm) | RRR | a (Å) | c (Å) |
|---|---|---|---|---|---|---|---|---|
| L-0 | 0 | 39.1 | 0.2 | 0.7 | 9.5 | 15 | 3.084 | 3.522 |
| L-1 | $1.0\cdot10^{15}$ | 39.1 | 0.2 | 0.7 | | 15 | | |
| L-2 | $1.0\cdot10^{16}$ | 39.1 | 0.2 | 0.7 | | 15 | | |
| L-3 | $1.0\cdot10^{17}$ | 39.1 | 0.2 | 2.0 | 14 | 8.0 | 3.086 | 3.526 |
| L-4 | $1.0\cdot10^{18}$ | 37.0 | 0.2 | 7.7 | 12 | 2.5 | 3.086 | 3.527 |
| P-0 | 0 | 39.1 | 0.2 | 1.6 | 16 | 11 | 3.084 | 3.519 |
| P-1 | $1.0\cdot10^{17}$ | 38.9 | 0.3 | 2.4 | 14 | 7.0 | 3.083 | 3.524 |
| P-2 | $6.0\cdot10^{17}$ | 37.7 | 0.2 | 6.5 | 13 | 3.0 | 3.085 | 3.529 |
| P-3 | $7.6\cdot10^{17}$ | 35.9 | 0.3 | 16 | 15 | 2.0 | 3.083 | 3.525 |
| P-3.5 | $2.0\cdot10^{18}$ | 33.3 | 0.3 | 26 | 14 | 1.6 | 3.088 | 3.537 |
| P-3.7 | $5.5\cdot10^{18}$ | 27.3 | 1.0 | 41 | 10 | 1.3 | 3.088 | 3.538 |
| P-4 | $1.0\cdot10^{19}$ | 23.8 | 0.9 | 64 | 13 | 1.2 | 3.088 | 3.549 |
| P-5 | $3.9\cdot10^{19}$ | 11.7 | 0.7 | 124 | 15 | 1.1 | 3.095 | 3.558 |
| P-6 | $1.4\cdot10^{20}$ | 9.1 | 0.3 | 130 | 12 | 1.1 | 3.093 | 3.558 |
| P-Cd1 | $1.6\cdot10^{18}$ | 39.0 | 0.2 | 2.1 | 15 | 8.3 | 3.084 | 3.521 |
| P-Cd2 | $5.6\cdot10^{18}$ | 38.9 | 0.2 | 1.9 | 11 | 6.7 | 3.086 | 3.526 |

**Table 2.** Fit parameters obtained by the $H_{c2}$ model for less irradiated samples of the P-series: $\rho_{0Mis.}$, resistivity measured at 40 K; $\eta$, ratio between diffusivities $D_\pi$ and $D_\sigma^{ab}$; $D_\sigma^{ab}$, $\sigma$-band in-plane diffusivity; $D_\pi$, $\pi$-band diffusivity; $\rho_{Calc.}$, resistivity calculated by $D_\pi$ and $D_\sigma^{ab}$; $H_{c2}(0)$, upper critical field at 0 K estimated by the model.

| Sample | $\rho_{0\ Mis.}$ | $\eta=(D_\pi/D_\sigma)$ | $D_\pi(\cdot10^{-4})$ | $D_\sigma(\cdot10^{-4})$ | $\rho_{Calc.}$ | $H_{c2}(0)$ |
|---|---|---|---|---|---|---|
| P-1 | 2.4 | 9.5 | 61 | 6.4 | 7 | 16.9 |
| P-2 | 6.5 | 8.6 | 43 | 4.9 | 10 | 21.0 |
| P-3 | 16 | 6.0 | 19 | 3.2 | 22 | 26.5 |
| P-3.5 | 26 | 5.7 | 11 | 2.0 | 37 | 30.2 |

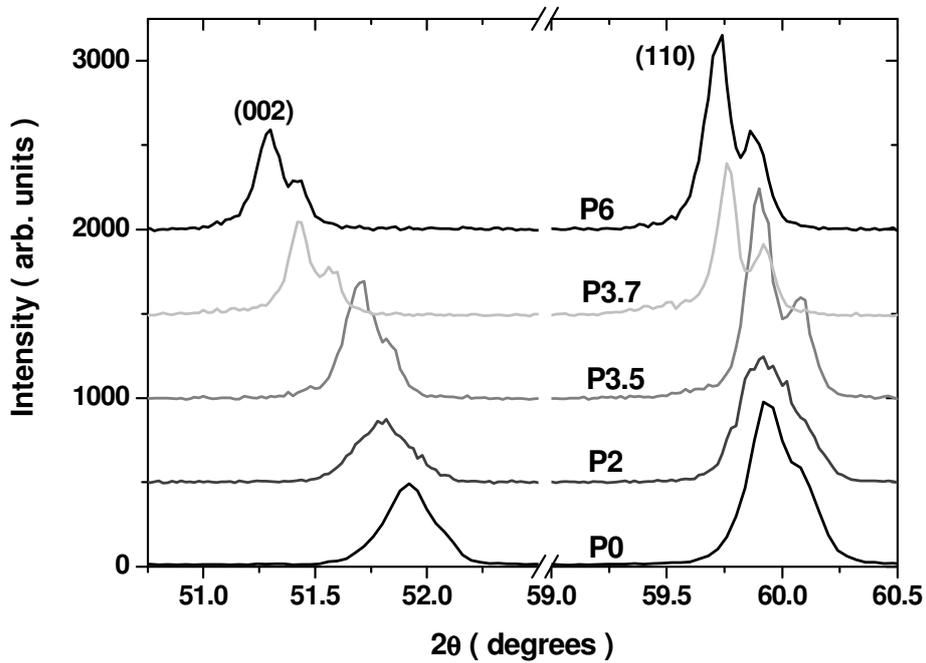

**Figure 1.** X-Ray diffraction patterns on (002) and (110) reflections for unirradiated (P0) and irradiated (P2, P3.5, P3.7, P6) samples.

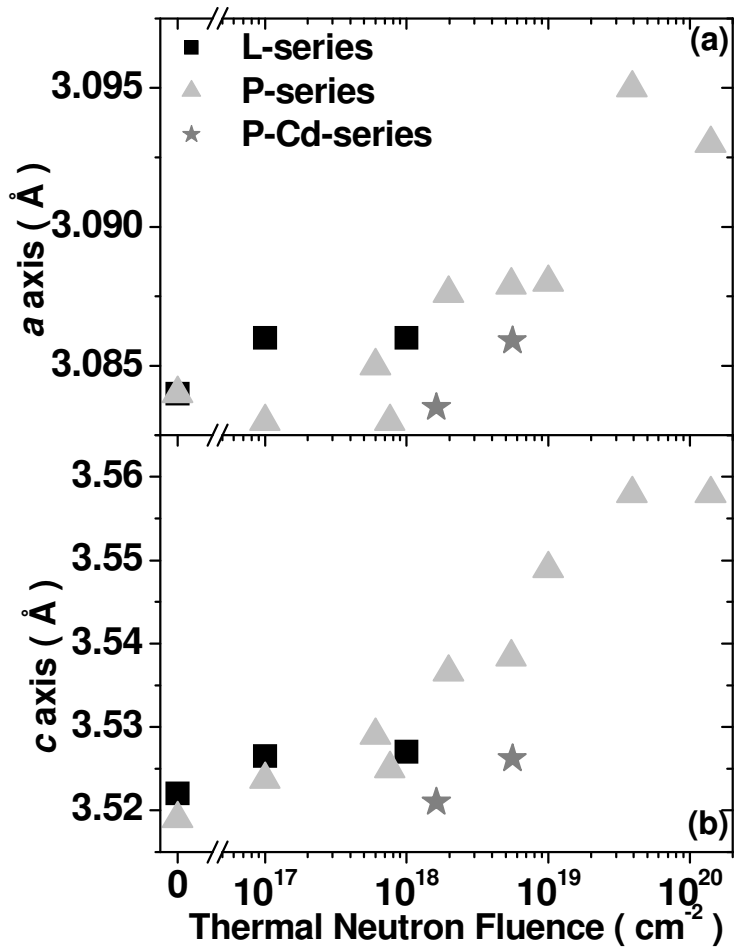

**Figure 2.** $a$ and $c$ axes as a function of thermal neutron fluence for L- (black squares), P- (grey triangles) and P-Cd-series (grey stars).

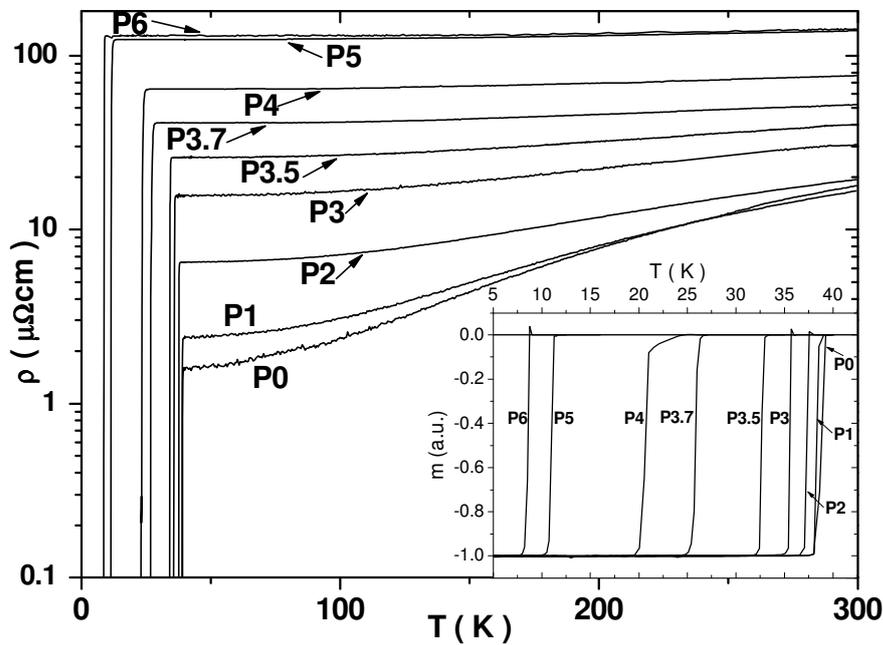

**Figure 3.** Resistivity curves as a function of temperature for the P-series; in the inset, the magnetic superconducting transitions are shown as measured in the SQUID.

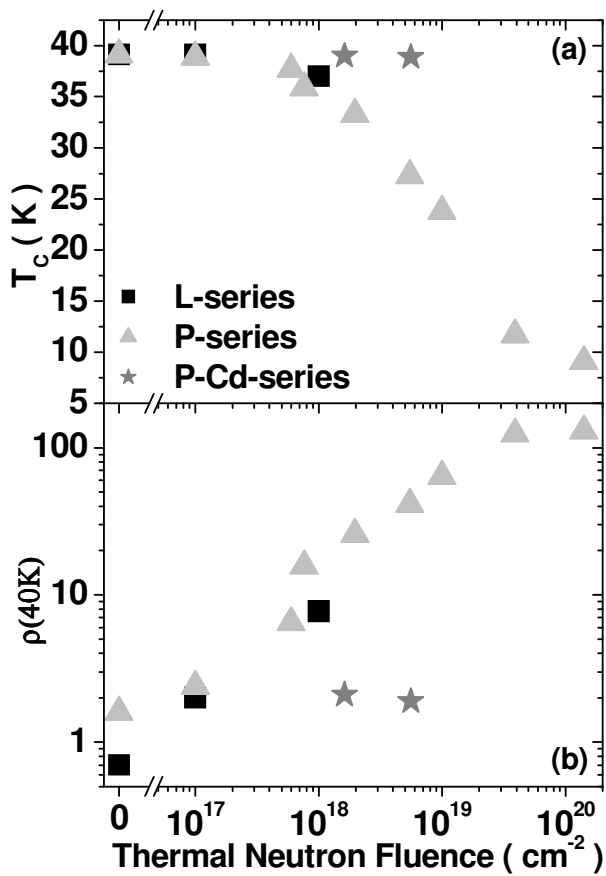

**Figure 4.** (a) Critical temperature $T_c$ and (b) resistivity at 40 K, $\rho(40K)$ as a function of thermal neutron fluence.

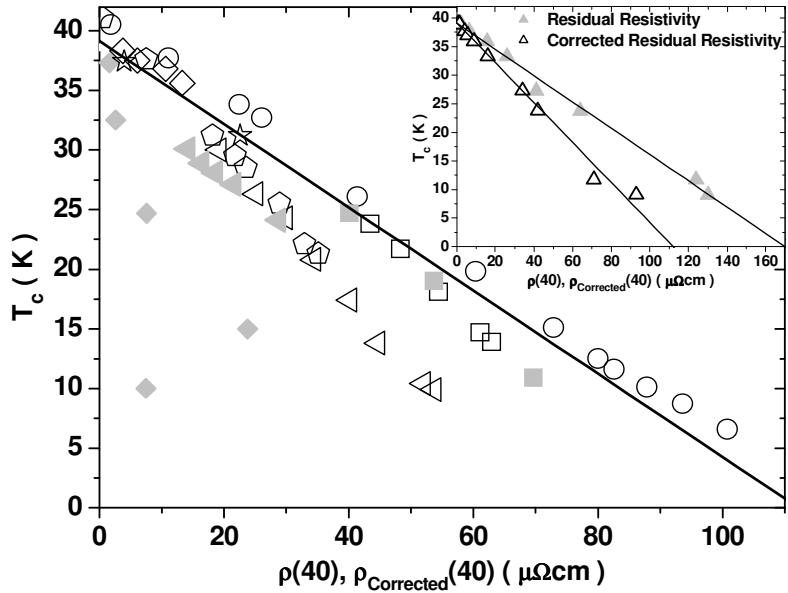

**Figure 5.** Critical temperature as a function of residual resistivity from various papers: ref.13, (○,⌾); ref.14, before (□,◁) and after (■,◀) annealing respectively; ref.27, (☆); ref.28, (◇); ref.19, (◆); the line represents the data of the present work. In the inset, critical temperature versus residual resistivity (full symbols) and versus residual resistivity corrected by Rowell's criterion (open symbols) for all the samples. The line, that represents the best fit of the open symbols, is also reported in the main panel.

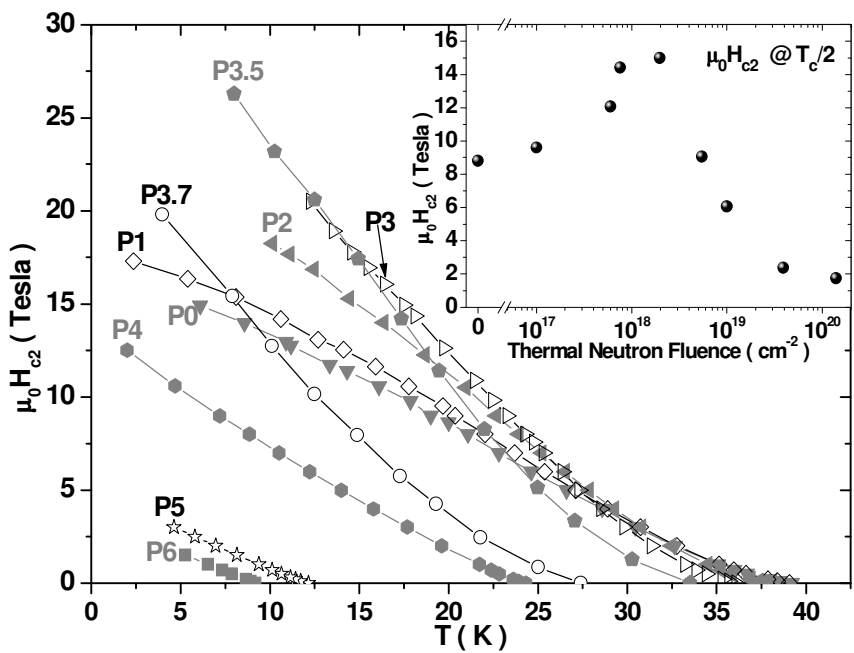

**Figure 6.** Upper critical field as a function of temperature estimated at 90% of the resistive transition. The inset shows critical field behaviour versus fluence at $T_c/2$ for the P-series.

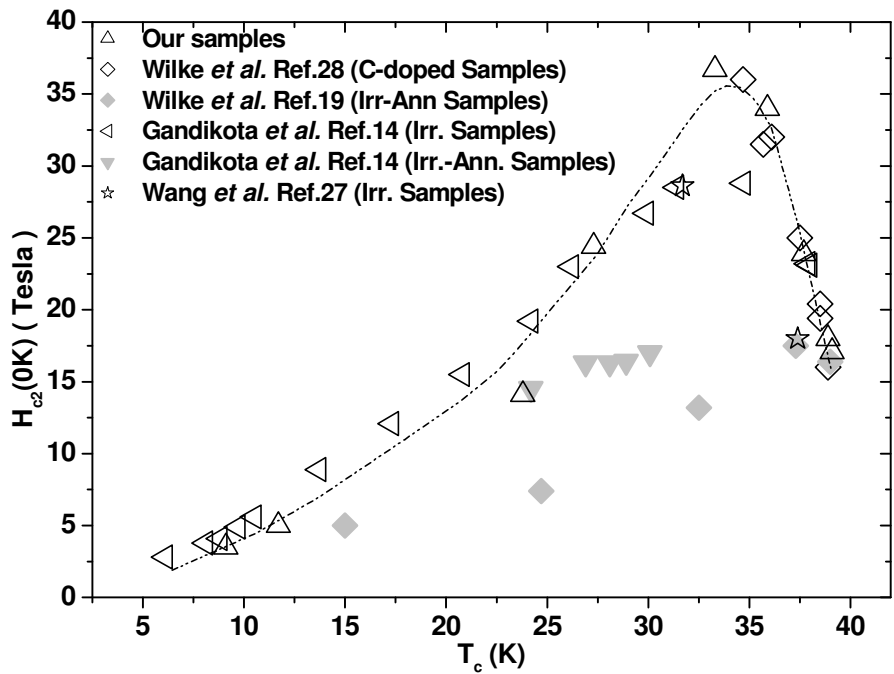

**Figure 7.** Upper critical field as a function of critical temperature in irradiated or C-doped samples (open symbols) and in irradiated and then annealed samples (full symbols); the line is only a guide for eyes.

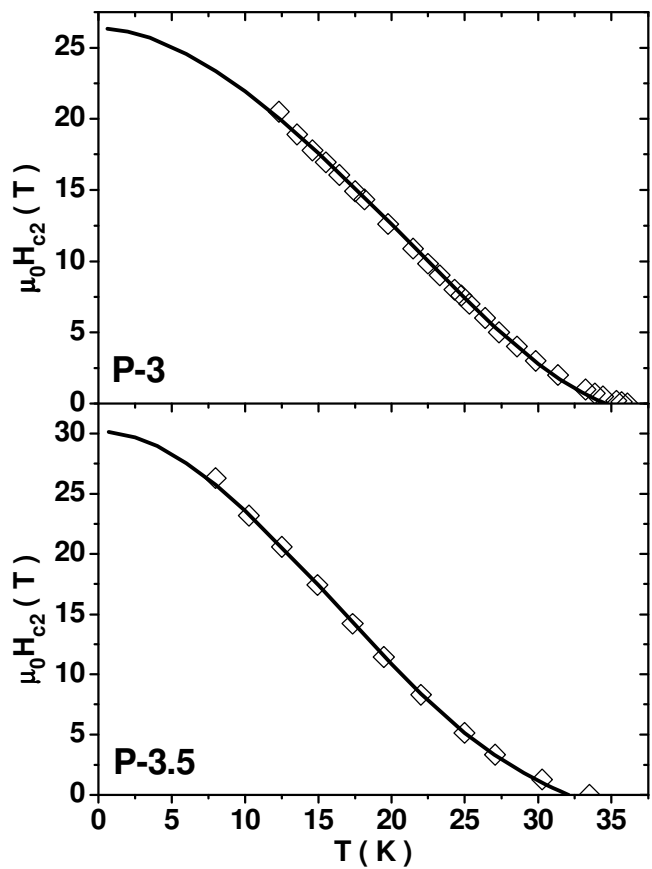

**Figure 8.** Upper critical field data (open symbols) and $H_{c2}$ fit curves (solid lines) for P3 and P3.5 samples.

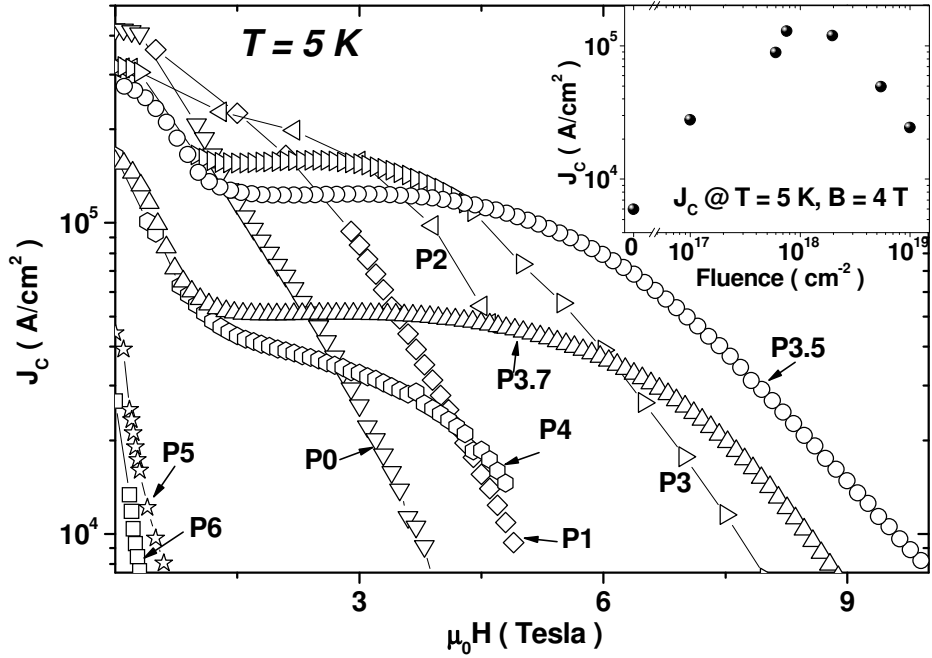

**Figure 9.** In the main panel, critical current density at 5 K as a function of magnetic field and as a function of fluence at 4 T, in the inset.

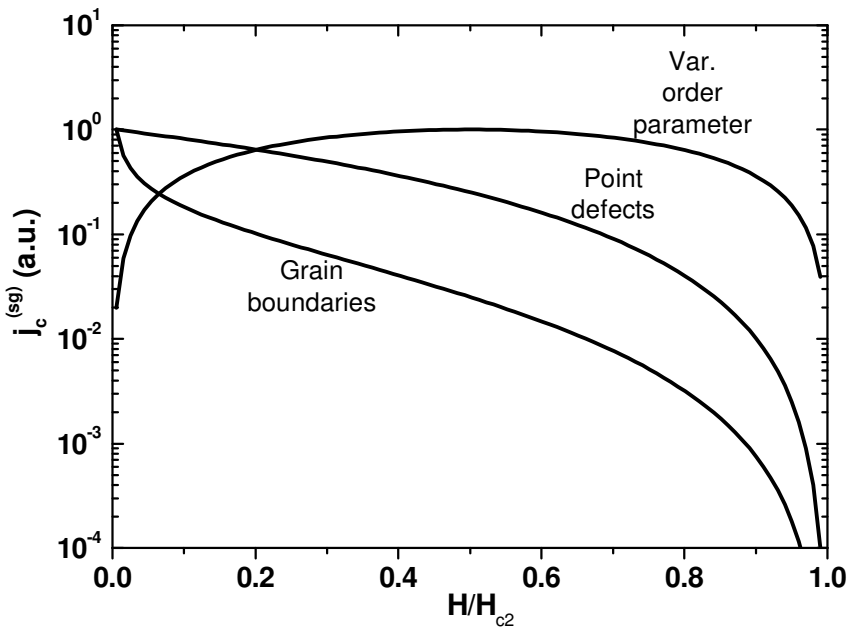

**Figure 10.** Single grain critical current density from three different pinning mechanisms (grain boundaries, point defects and variations of the superconducting order parameter) as a function of the reduced field $H/H_{c2}$. The curves are normalized to the respective maximum values.

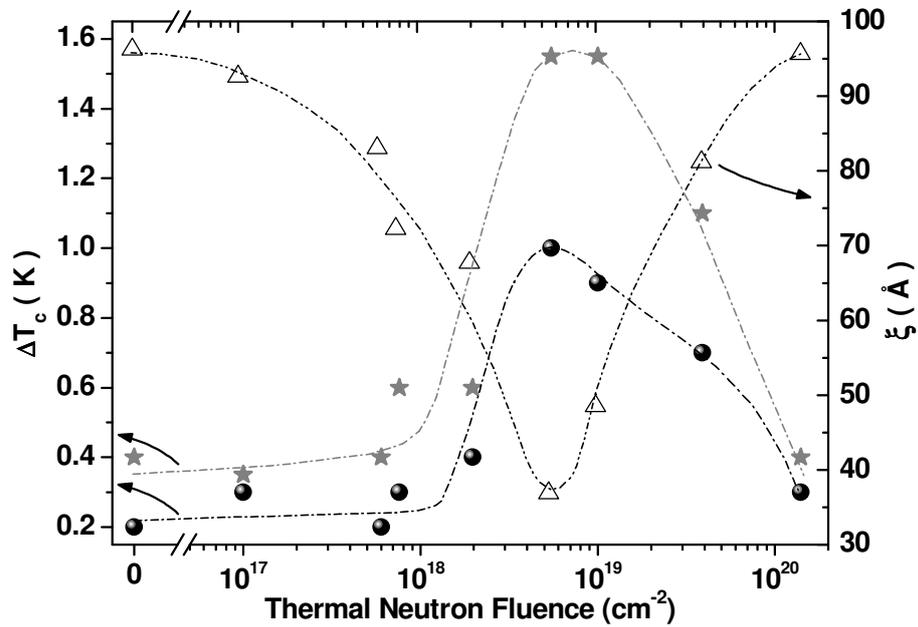

**Figure 11.** Transition width $\Delta T_c$ ($T_{90\%}$ - $T_{10\%}$, black circles, $T_{98\%}$ - $T_{2\%}$, grey stars) and coherence length $\xi$ (open symbols) as a function of thermal neutron fluence: the lines are guides for eyes.

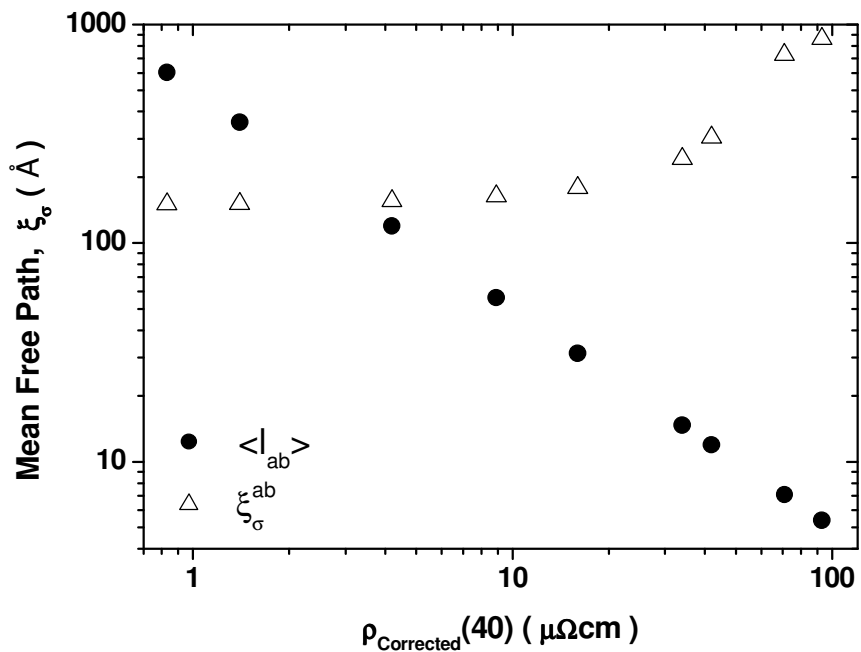

**Figure 12.** In-plane BCS coherence length $\xi_0$ for the σ-band (open symbols) and in-plane mean free path (solid symbols) as a function of corrected residual resistivity.